\documentclass[aps,prl,twocolumn,showpacs,superscriptaddress,groupedaddress]{revtex4}

\usepackage{bbm}

\usepackage{graphicx}
\usepackage{dcolumn}
\usepackage{amsmath}
\usepackage{bm}
\usepackage{color}
\usepackage{mathrsfs}
\usepackage{epstopdf}
\usepackage{amsmath}
\usepackage{amssymb}
\usepackage{subfigure}
\usepackage{enumitem}
\usepackage{multirow}
\usepackage{exscale}
\usepackage{relsize}

\newcommand{\be}{\begin{equation}}
\newcommand{\ee}{\end{equation}}
\newcommand{\bey}{\begin{eqnarray}}
\newcommand{\eey}{\end{eqnarray}}
\newcommand{\bw}{\begin{widetext}}
\newcommand{\ew}{\end{widetext}}

\newcommand{\ba}{\begin{array}}
\newcommand{\ea}{\end{array}}
\newcommand{\bi}{\begin{itemize}}
\newcommand{\ei}{\end{itemize}}
\newcommand{\bem}{\begin{enumerate}}
\newcommand{\eem}{\end{enumerate}}

\begin{document}

 \title {Nonequilibrium Green's function's approach to the calculation of work statistics
 }

\author{Zhaoyu Fei}
\affiliation{School of Physics, Peking University, Beijing 100871, China}

\author{H. T. Quan} \email[Email: ]{htquan@pku.edu.cn}
\affiliation{School of Physics, Peking University, Beijing 100871, China}
\affiliation{Collaborative Innovation Center of Quantum Matter, Beijing 100871, China}
\affiliation{Frontiers Science Center for Nano-optoelectronics, Peking University, Beijing, 100871, China}

 \date{\today}

\begin{abstract}
   The calculation of work distributions in a quantum many-body system is of significant importance and also of formidable difficulty in the field of nonequilibrium quantum statistical mechanics.
   To solve this problem, inspired by the Schwinger-Keldysh formalism, we propose the contour-integral formulation for work statistics. Based on this contour integral, we show how to do the perturbation expansion of the characteristic function of work (CFW) and obtain the approximate expression of the CFW to the second order of the work parameter for an arbitrary system under a perturbative protocol. We also demonstrate the validity of fluctuation theorems by utilizing the Kubo-Martin-Schwinger condition. Finally, we use noninteracting identical particles in a forced harmonic potential as an example to demonstrate the powerfulness of our approach.
 \end{abstract}


 \maketitle

\textit{Introduction}.---In the past 25 years or so, the development of stochastic thermodynamics and the discovery of fluctuation theorems have revolutionized our understanding about nonequilibrium thermodynamics~\cite{st2010,eq2011,st2012,non2013}. In these studies, a key quantity is the probability distribution of work in an arbitrary nonequilibrium process, which encodes essential information about the nonequilibrium process analogous to the partition function encoding essential information about an equilibrium state~\cite{ge2012,th2018,as2015,wor2019}.
For a closed quantum system, the trajectory work is defined as the difference between the results of the projective measurements over the system's energy before and after the driving protocol~\cite{aq2000,ja2000,flu2007}. Accordingly, the characteristic function of work (CFW, the Fourier transform of the work distribution $P(w)$) reads~\cite{flu2007}
\be
\label{e1}
\chi(v)=\int\mathrm dw P(w)e^{ivw}=\mathrm{Tr}[\hat{U}^{\dag}(t)e^{iv\hat H(t)}\hat{U}(t)e^{-iv\hat H(0)}\hat\rho],
\ee
where $\hat \rho$ denotes the initial state, $\hat H(0)$ and $\hat H(t)$ denote the Hamiltonians before and after the driving protocol and $\hat{U}(s)$ denotes the time-evolution operator corresponding to a time-dependent Hamiltonian $\hat H(s),s\in[0,t]$. 
The CFW is a powerful tool to study the nonequlilibrium physics of a quantum system since it appears not only in stochastic thermodynamics, but also in Loschmidt echoes~\cite{st2008,de2006}, Kibble-Zurek mechanism~\cite{fzy2020}, dynamical quantum phase transitions~\cite{dy2013,qu2016} and many other fields. Hence, to efficiently calculate the CFW becomes one of the most important problems in this field.

Nevertheless, it is usually a very challenging task to calculate the CFW for an arbitrary nonequilibrium protocol, especially for quantum many-body systems, due to the complicated nonequilibrium dynamics. In the literature, there are a few results about the CFW, but mostly focusing on special models and are studied case by case~\cite{sta2008,qu2017,sta2019,no2008,wo2013,wor2019,fu2019,wo2019,qu2019,em2012,ja2018}. For example, in Refs.~\cite{ja2018,wor2019}, the perturbation expansion is applied to the calculation of the work distrbutions of a quantum scalar field for perturbative protocols. For quantum systems described by quadratic Hamiltonians, Ref.~\cite{gr2019} proposed a general method for solving the CFW under an arbitrary driving protocol by utilizing the group-representation theory. Nevertheless, for a general model beyond the quadratic Hamiltonian, no efficient ways to solve the CFW have been reported so far.

In this letter, in order to address the above problem, we propose the nonequilibrium Green's function's approach to the calculation of the CFW. Based on the Schwinger-Keldysh formalism~\cite{di1965,no2013}, nonequilibrium Green's functions provide a useful framework to handle problems of  time-dependent Hamiltonians. For example, it is a standard tool in deriving Landauer formula in quantum transport~\cite{la1992}. Also, it has been applied to the calculation of the full counting statistics of heat~\cite{fl2007}. Inspired by this formalism, we propose the contour for work statistics and define the work functional along the modified contour. In this way, we are able to calculate the CFW of an arbitrary system for a perturbative protocol by the perturbation expansion. Also, to the second order of the expansion, we obtain the general expression of the CFW and demonstrate the fluctuation theorems by utilizing the Kubo-Martin-Schwinger condition~\cite{no2013}. 

We also notice that Refs.~\cite{no2018,sy2019} discussed the work statistics and fluctuation theorems based on the Schwinger-Keldysh formalism. Different from our modified contour for work statistics, they defined the modified Hamiltonian on the usual Schwinger-Keldysh contour. In contrast to their method, where the explicit expression of the modified Hamiltonian is usually difficult to obtain, the correlation functions in our paper (see below) can be more readily calculated, which significantly simplifies the calculation of the CFW.

\begin{figure*}
\includegraphics[width=\textwidth]{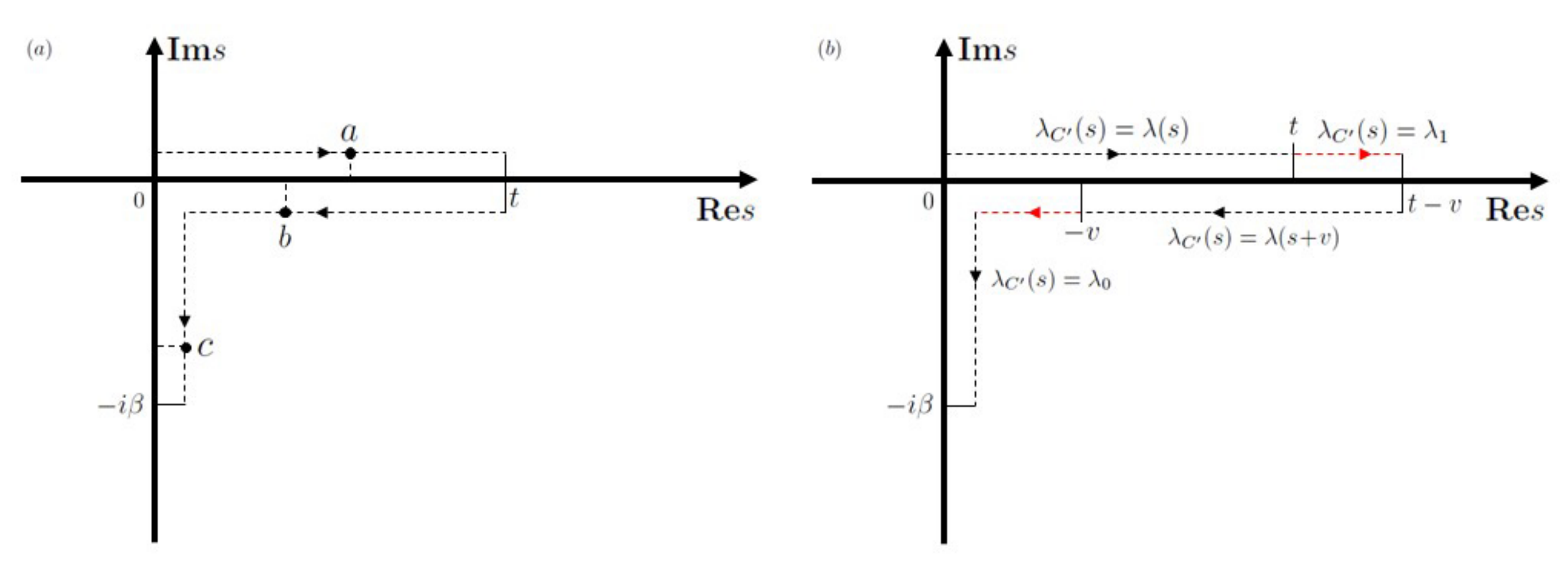}
\caption{Complex plane of time $s$. (a) The Schwinger-Keldysh contour $C$. Contour ordering: $a<b<c$. (b) The contour for work statistics $C'$ ($\lambda_{C'}(s)=\lambda_0$ in the last two branches). The red lines denote the exponential operators in Eq.~(\ref{e1}). We have assumed $v<0$ in (b), which does not influence the calculation of the CFW.}\label{fig1}
\end{figure*}

\textit{From the Schwinger–Keldysh contour to the contour for work statistics}.---For a time-dependent quantum system $\hat H(s)=\hat H_0+\lambda(s)\hat H_1$, $s\in[0,t]$ with the work parameter $\lambda(s)$ and the canonical initial state $\hat \rho=e^{-\beta\hat H(0)}/\mathrm{Tr}[e^{-\beta\hat H(0)}]$ ($\beta=(k_B T)^{-1}$ is the inverse temperature), the expectation value of an observable $\hat O$ at time $t$ is $\mathrm{Tr}[\hat\rho^I(t) \hat O^I(t)]$, where the time-dependent operators are in the interaction picture, $\hat O^I(t)=e^{\frac{i}{\hbar}\hat H_0 t}\hat O e^{-\frac{i}{\hbar}\hat H_0 t}$, $\hat \rho^I(t)= e^{\frac{i}{\hbar}\hat H_0 t}\hat U(t,0)\hat\rho\hat U(0,t)e^{-\frac{i}{\hbar}\hat H_0 t}$. In the Schwinger-Keldysh formalism, this quantity is related to a contour with three directed branches, called the Schwinger-Keldysh contour (see Fig.~\ref{fig1}a). Thus, the expectation value can be calculated by a contour integral $\mathrm{Tr}[\hat\rho^I(t) \hat O^I(t)]=\langle \mathcal{T}_C[\hat O^I(t)e^{-\frac{i}{\hbar}\int_C \mathrm{d}s\lambda(s)\hat H^I_1(s)}]\rangle/\langle \mathcal{T}_C[e^{-\frac{i}{\hbar}\int_{0}^{-i\hbar\beta}  \mathrm{d}s\lambda_0\hat H^I_1(s)}]\rangle$, where $\langle \cdot \rangle=\mathrm{Tr}[\cdot e^{-\beta\hat H_0}]/\mathrm{Tr}[e^{-\beta\hat H_0}]$, the integral is along the contour $C$ and $\mathcal{T}_C$ indicates ordering along the same contour (e.g., $a<b<c$ in Fig.~\ref{fig1}a)~\cite{di1965,no2013}.

Inspired by this formalism, we treat both the time-evolution operators and the exponential operators in Eq.~(\ref{e1}) as the directed branches of a modified contour $C'$ (see Fig.~\ref{fig1}b). Then the contour-integral formulation of the CFW reads
\be
\label{e2}
\chi(v)=\frac{\langle \mathcal{T}_{C'}[e^{-\frac{i}{\hbar}\int_{C'} \mathrm{d}s\lambda_{C'}(s)\hat H^I_1(s)}]\rangle}{\langle \mathcal{T}_{C'}[e^{-\frac{i}{\hbar}\int_{0}^{-i\hbar\beta} \mathrm{d}s\lambda_0\hat H^I_1(s)}]\rangle},
\ee
where the integral, $\mathcal{T}_{C'}$ and the work parameter $\lambda_{C'}(t)$ are all along the new contour $C'$. Hence, we call the new contour $C'$ the contour for work statistics, which is also consistent with the Ramsey interferometry~\cite{ex2013} and the work statistics in the path integral formalism~\cite{pa2018}. Moreover, Eq.~(\ref{e2}) can be rewritten as follows
 \begin{gather}
  \begin{split}
\chi(v)=&\frac{\langle \mathcal{T}_{C'}[e^{\frac{i}{\hbar}\int_{0}^{t}\mathrm ds\int_{0}^{\hbar v}\mathrm dr \dot{\lambda}(s)\hat{H}^I_1(s-r)}]\mathcal{T}_{C'}[e^{-\frac{i}{\hbar}\int_{0}^{-i\hbar\beta} \mathrm{d}s\lambda_0\hat H^I_1(s)}]\rangle}{\langle \mathcal{T}_{C'}[e^{-\frac{i}{\hbar}\int_{0}^{-i\hbar\beta}\mathrm{d}s\lambda_0\hat H^I_1(s)}]\rangle}\\
\equiv&\langle \mathcal{T}_{C'}[e^{iv\hat W}]\rangle'
  \end{split}
 \end{gather}
where $\dot{\lambda}(s)=\mathrm{d}\lambda(s)/\mathrm{d}s$ and $\langle\cdot\rangle'=\mathrm{Tr}\{\cdot\mathcal{T}_{C'}[e^{-\frac{i}{\hbar}\int_{0}^{-i\hbar\beta} \mathrm{d}s\lambda_0\hat H^I_1(s)}]\}/\mathrm{Tr}\{\mathcal{T}_{C'}[e^{-\frac{i}{\hbar}\int_{0}^{-i\hbar\beta} \mathrm{d}s\lambda_0\hat H^I_1(s)}]\}$. Here, we call $\hat W=\frac{1}{\hbar v}\int_{0}^{t}\mathrm ds\int_{0}^{\hbar v}\mathrm dr \dot{\lambda}(s)\hat{H}^I_1(s-r)$ the work functional (similar to the work functional defined in Ref.~\cite{pa2018}). In the classical limit ($\hbar\to 0$), the time-ordered operator $\mathcal{T}_{C'}$ disappears and the work functional $\hat W$ just corresponds to the classical trajectory work $W[x(s),p(s)]=\int_{0}^{t}\mathrm{d}s\dot{\lambda}(s)H_1(x(s),p(s),s)$~\cite{no1997}. However, this does not mean that work is an observable~\cite{flu2007}. Actually, the work functional $\hat W$ is  the combination of the operators in different branches of $C'$. Hence, it is nonsense to consider the eigenstates or eigenvalues of $\hat W$ due to $\mathcal{T}_{C'}$.

\textit{Calculating work statistics based on the perturbation expansion}.---The exponential operator in Eq.~(\ref{e2}) can be expanded as
\be
\label{e4}
\chi(v)=\frac{1+\sum_{n=1}^{\infty} \left(\prod_{l=1}^{n}\int_{C'}\mathrm d\bar{s}_l\right) G(s_1,\cdots,s_n)}{1+\sum_{n=1}^{\infty}\left(\prod_{l=1}^{n}\int_{0}^{-i\hbar\beta} \mathrm d\bar{s}_l\right)G(s_1,\cdots,s_n)},
\ee
where $\mathrm d\bar{s}_l=\mathrm{d}s_l\lambda_{C'}(s_l)\theta_{C'}(s_l-s_{l+1})$ is an abbreviation,
\be
G(s_1,\cdots,s_n)=\left(\frac{-i}{\hbar}\right)^n\langle\hat H^I_1(s_1)\cdots\hat H^I_1(s_n)\rangle
\ee
is the $n$-point correlation function, $\theta_{C'}(s-s')$ is the contour step function~\cite{no2013} and we set $\theta_{C'}(s_n-s_{n+1})\equiv1$. A more convenient notion is the series of the logarithm of $\chi(v)$, called the cumulant CFW (see supplemental material),
\be
\label{e6}
\ln\chi(v)=\sum_{n=1}^{\infty}\left(\prod_{l=1}^{n}\int_{C'} \mathrm d\bar{s}_l- \prod_{l=1}^{n}\int_{0}^{-i\hbar\beta}\mathrm d\bar{s}_l\right)G_c(s_1,\cdots,s_n),
\ee
where
\be
\label{e7}
G_c(s_1,\cdots,s_n)=\left(\frac{-i}{\hbar}\right)^n\langle\hat H^I_1(s_1)\cdots\hat H^I_1(s_n)\rangle_c
\ee
is the $n$-point cumulant correlation function (also called Ursell function)~\cite{qu1987,th1952}.
For the perturbative driving protocol $\lambda(s)$, Eqs.~(\ref{e4}, \ref{e6}) are the perturbation expansion of the work statistics.
Usually, we are able to calculate the correlation functions by Wick's theorem and Feynman diagrams~\cite{gr2006}. Here in $G_c(s_1,\cdots,s_n)$, only connected diagrams are included.

After a straightforward calculation (see supplemental material), to the second order of $\lambda(s)$, we obtain the approximate expression of the perturbation expansion of $\ln\chi(v)$ for a perturbative protocol
 \begin{gather}
  \begin{split}
  \label{e8}
\ln\chi(v)= &iv(\lambda_1-\lambda_0)\langle\hat{H}_1\rangle_c+\int_{-\infty}^{\infty}\frac{\mathrm{d}\omega}{2\pi}\frac{1-e^{i\omega \hbar v}}{\omega^2}A(\omega)G^>_c(\omega)\\
&+i\hbar v(\lambda_1^2-\lambda_0^2)\int_{-\infty}^{\infty}\frac{\mathrm{d}\omega}{2\pi}\frac{G^>_c(\omega)}{\omega}+O(\lambda(s)^3).
  \end{split}
 \end{gather}
Here, the first term on the RHS of Eq.~(\ref{e8}) represents the average work done to the first order of $\lambda(s)$.
The second term is called the ``speed'' term since $A(\omega)$ depends on $\dot{\lambda}(s)$ by
\be
A(\omega)=\left|\int_{0}^{t}\mathrm ds \dot{\lambda}(s)e^{i\omega s}\right|^2.
\ee
It encodes the information about the nonequilibrium protocols $\lambda(s)$.
The third term is called the ``boundary'' term since only the initial and the final value of $\lambda(s)$ appear in this term. Meanwhile, the information about the Hamiltonians $\hat H_0$ and $\hat H_1$ is encoded in the cumulant greater correlation function $G^>_c(\omega)$
\be
\label{e10}
G^>_c(\omega)=\int_{-\infty}^{\infty}\mathrm ds G_c^>(s)e^{i\omega s}, G^>_c(s)=\left(\frac{-i}{\hbar}\right)^2\langle \hat H^I_1(s)\hat H^I_1(0)\rangle_c.
\ee
We would like to emphasize that Eq.~(\ref{e8}) is one of the main results in our paper. It is a general result of the work statistics because it is valid for arbitrary $\hat H_0$ and $\hat H_1$, and also arbitrary perturbative protocols $\lambda(s)$.

In the following, we analyze the properties of the CFW based on our results (Eq.~(\ref{e8})). Above all, the CFW satisfies the normalization condition ($\ln \chi (0)=0$). As for the fluctuation theorems, let us first introduce the backward process of $\hat{H}(s)$:
$\hat H_B(s)=\hat H_0+\lambda(t-s)\hat H_1,\hat \rho_{B}=e^{-\beta\hat H_B(0)}/\mathrm{Tr}[e^{-\beta\hat H_B(0)}]$~\cite{foot1}. Then, the perturbation expansion of $\ln\chi_B(v)$ to the second order of $\lambda(s)$ can be written as
 \begin{gather}
  \begin{split}
  \label{e11}
&\ln\chi_B(v)=\\
&-iv(\lambda_1-\lambda_0)\langle\hat{H}_1\rangle_c+\int_{-\infty}^{\infty}\frac{\mathrm{d}\omega}{2\pi}\frac{1-e^{i\omega \hbar v}}{\omega^2}A(\omega)G^>_c(\omega)\\
&-i\hbar v(\lambda_1^2-\lambda_0^2)\int_{-\infty}^{\infty}\frac{\mathrm{d}\omega}{2\pi}\frac{G^>_c(\omega)}{\omega}+O(\lambda(s)^3).
  \end{split}
 \end{gather}
Thus according to the Kubo-Martin-Schwinger condition~\cite{no2013}, $G^>_c(s-i\hbar\beta)=G^>_c(-s),e^{-\beta\hbar\omega}G^>_c(\omega)=G^>_c(-\omega)$, and $A(\omega)$ is an even function, from Eqs.~(\ref{e8}, \ref{e11}) we obtain the following relation
 \begin{gather}
  \begin{split}
  \label{e12}
&-k_BT\ln\frac{\chi(-v+i\beta)}{\chi_B(v)}=(\lambda_1-\lambda_0)\langle\hat{H}_1\rangle_c \\ &+\hbar(\lambda_1^2-\lambda_0^2)\int_{-\infty}^{\infty}\frac{\mathrm{d}\omega}{2\pi}\frac{G^>_c(\omega)}{\omega}+O(\lambda(s)^3),
  \end{split}
 \end{gather}
where the first and second terms on the RHS of Eq.~(\ref{e12}) are exactly the perturbation expansion of the free energy difference $\Delta F=F_{\lambda_1}-F_{\lambda_0}=-k_BT\ln(\mathrm{Tr}[e^{-\beta\hat H(t)}]/\mathrm{Tr}[e^{-\beta\hat H(0)}])$ to the first and the second order of $\lambda(s)$ respectively~\cite{on1951,th1952}. After taking the inverse Fourier transform of Eq.~(\ref{e12}), we obtain the relation between the work distributions  of the forward and the backward processes ($P(w)$ and $P_B(w)$) to the second order of $\lambda(s)$
\be
\label{e13}
\frac{P(w)}{P_B(-w)}=e^{\beta(w-\Delta F)}+O(\lambda(s)^3),
\ee
which is nothing but the Crooks fluctuation theorem~\cite{th2009,en1999} to the second-order perturbation expansion of $\lambda(s)$.
As a result, Jarzynski equality~\cite{no1997} to the second-order perturbation expansion can be obtained as a straightforward corollary of Eq.~(\ref{e13})~\cite{eq2011}.

\textit{Example: noninteracting identical particles in a forced harmonic potential}.---To calculate the CFW of a quantum many-body system in an arbitrary nonequilibrium process is an extremely cumbersome task. However, for those perturbative driving protocols, our method based on the nonequilibrium Green's function provides a unified and powerful tool to solve this long-standing problem. We demonstrate our results by considering the following time-dependent Hamiltonian
 \begin{gather}
  \begin{split}
  \hat H_0=&\sum_{i=1}^{N}\frac{\hat {p}_i^2}{2m}+\frac{m}{2}(\omega_x^2\hat {x}_i^2+\omega_y^2\hat {y}_i^2+\omega_z^2\hat {z}_i^2),\\
  \hat H_1=&\omega_z\sqrt{2m}\sum_{i=1}^{N}\hat {z}_i,
  \end{split}
 \end{gather}
which describes noninteracting spinless identical particles in a 3-dimensional harmonic potential driven along the $z$ direction with the particle mass $m$, the total particle number $N$ and the frequencies along three directions $\omega_x,\omega_y,\omega_z$. For simplicity, we choose $\omega_x\sim \omega_y\sim \omega_z$. This system is a well-known physical model in statistical physics (e.g., see Ref.~\cite{st2011}). For bosons, it is a good model to study Bose-Einstein condensation in non-interacting trapped gases~\cite{th1999,st2011}. The transition temperature $k_BT_c$ equals $\hbar\omega_g\left[N/\zeta(3)\right]^{1/3}$, where $\omega_g=(\omega_x\omega_y\omega_z)^{1/3}$ and $\zeta(s)$ is the Riemann zeta function. Moreover, the proper thermodynamic limit for these systems is obtained by letting $N\to \infty$ and $\omega_{g}\to 0$, while keeping the product $N\omega_{g}^3$ as a constant. We would like to emphasize that $\lambda(s)$ does not depend on $N$. As a result, for various $N$ and the same $\lambda(s)$, we always  have $\Delta F/F_{\lambda=0}\sim O(N^0)$ for the canonical ensemble.

After the second quantization, $\hat H_0$ and $\hat H_1$ read
 \begin{gather}
  \begin{split}
  \hat H_0=&\sum_{\mathbf k}\varepsilon_{\mathbf k}\hat a_{\mathbf k}^\dag\hat a_{\mathbf k}\\
  \hat H_1=&\sqrt{\hbar \omega_z}\sum_{\mathbf k}\sqrt{k_z+1}(\hat a_{\mathbf k}^\dag\hat a_{\tilde{\mathbf k}}+\hat a_{\tilde{\mathbf k}}^\dag\hat a_{\mathbf k}),
  \end{split}
 \end{gather}
where $\mathbf k=(k_x,k_y,k_z)\in{\mathbb N}^3$, $\tilde{\mathbf k}=(k_x,k_y,k_z+1)$, and $\varepsilon_{\mathbf k}=\hbar(k_x\omega_x+k_y\omega_y+k_z\omega_z)+\varepsilon_{0}$ is the single-particle-state energy, $\varepsilon_{0}=\hbar(\omega_x+\omega_y+\omega_z)/2$. For later convenience, let us introduce the following notations: the total particle number operator $\hat N=\sum_{\mathbf k}\hat n_{\mathbf k}=\sum_{\mathbf k}\hat a_{\mathbf k}^\dag\hat a_{\mathbf k}$ with its eigenstates $\left|\left\{n_{\mathbf k}\right\}\right\rangle=\otimes_{\mathbf k}\left|n_{\mathbf k}\right\rangle$;
for the canonical ensemble, the density matrix $\hat \rho_N=\delta(\hat N-N)e^{-\beta\hat H_0}/Z_N$, where $Z_N=\mathrm{Tr}[\delta(\hat N-N)e^{-\beta\hat H_0}]$,
$\left\langle\left\{n_{\mathbf k}\right\}\right| \delta(\hat N-N)|\{n'_{\mathbf k}\}\rangle=\delta_{\left\{n_{\mathbf k}\right\},\{n'_{\mathbf k}\}}\delta_{N,\sum_{\mathbf k}n_{\mathbf k}}$. Here $\delta_{\cdot,\cdot}$ is the Kronecker delta function; the mean occupation number in the canonical ensemble $\overline{n}_{\mathbf k}(N)=\mathrm{Tr}[\hat \rho_N \hat n_{\mathbf k}]$; for the grand canonical ensemble, the density matrix $\hat \rho_{\mu}=e^{-\beta(\hat H_0-\mu\hat N)}/Z_{\mu}$, where $Z_{\mu}=\mathrm{Tr}[e^{-\beta(\hat H_0-\mu\hat N)}]=\prod_{\mathbf k}(1-\gamma\alpha e^{-\beta\varepsilon_{\mathbf k}})^{-\gamma}$, $\gamma=1,-1$ for bosons and fermions respectively; the fugacity $\alpha=e^{\beta\mu}$; the mean occupation number in the grand canonical ensemble $\overline{n}_{\mathbf k}(\mu)=\mathrm{Tr}[\hat \rho_\mu \hat n_{\mathbf k}]=1/[e^{\beta(\varepsilon_{\mathbf k}-\mu)}-\gamma]$. In addition, quantities in these two ensembles are related by the fugacity expansions~\cite{st2011}
 \begin{gather}
  \begin{split}
  \label{e17}
  Z_\mu=&1+\sum_{N=1}^{\infty}\alpha^NZ_N\\
  Z_{\mu}\overline{n}_{\mathbf k}(\mu)=&\sum_{N=1}^{\infty}\alpha^NZ_N\overline{n}_{\mathbf k}(N).
  \end{split}
 \end{gather}
Similarly, we obtain the relation between the CFWs in these two ensembles
 \begin{gather}
  \begin{split}
  Z_{\mu}\chi_{\mu}(v)=&1+\sum_{N=1}^{\infty}\alpha^NZ_N\chi_N(v).
  \end{split}
 \end{gather}
Thus according to Eq.~(\ref{e10}) and Wick's theorem in the grand canonical ensemble, we obtain the cumulant greater correlation function in the grand canonical ensemble
 \begin{gather}
  \begin{split}
  \label{e19}
&G^>_{\mu}(\omega)=-\frac{2\pi\omega_z}{\hbar}\sum_{\mathbf k}(k_z+1)\{\delta(\omega-\omega_z)[\overline {n}_{\mathbf k}(\mu)\\
&+\gamma \overline {n}_{\mathbf k}(\mu)\overline {n}_{\tilde{\mathbf k}}(\mu)]+\delta(\omega+\omega_z)[\overline {n}_{\tilde{\mathbf k}}(\mu)+\gamma\overline {n}_{\mathbf k}(\mu)\overline {n}_{\tilde{\mathbf k}}(\mu)]\}.
  \end{split}
 \end{gather}
Then from Eqs.~(\ref{e17}, \ref{e19}), we obtain the cumulant greater correlation function in the canonical ensemble
 \begin{gather}
  \begin{split}
  \label{e18}
G^>_{N}(\omega)=&-\frac{2\pi\omega_z}{\hbar}\sum_{\mathbf k}(k_z+1)\{\delta(\omega-\omega_z)[\overline {n}_{\mathbf k}(N)\\
&+\gamma \xi_{\mathbf k}(N)]+\delta(\omega+\omega_z)[\overline {n}_{\tilde{\mathbf k}}(N)+\gamma \xi_{\mathbf k}(N)]\},
  \end{split}
 \end{gather}
where we have defined $\xi_{\mathbf k}(N)$ as
\be
\label{e20}
Z_{\mu}\overline {n}_{\mathbf k}(\mu)\overline {n}_{\tilde{\mathbf k}}(\mu)=\sum_{N=1}^{\infty}\alpha^NZ_N\xi_{\mathbf k}(N).
\ee
Finally, substituting Eqs.~(\ref{e19}, \ref{e18}) in Eq.~(\ref{e8}) and considering $\langle\hat{H}_1\rangle_c=0$, we obtain the perturbation expansion of the cumulant CFW to the second order of $\lambda(s)$ with the canonical and the grand canonical initial state respectively
 \begin{gather}
  \begin{split}
  \label{e22}
&\ln \chi_N(v)\approx-4\sum_{\mathbf k}[\gamma (k_z+1)\xi_{\mathbf k}(N)+k_z\overline {n}_{\mathbf k}(N)]\times\\
              &\frac{\sin^2(v\hbar\omega_z/2)}{\hbar\omega_z}A(\omega_z)+N\left[-iv(\lambda_1^2-\lambda_0^2)+\frac{e^{iv\hbar\omega_z}-1}{\hbar\omega_z}A(\omega_z)\right],
  \end{split}
 \end{gather}
  \begin{gather}
  \begin{split}
  \label{e23}
&\ln \chi_{\mu}(v)\approx-4\sum_{\mathbf k}[\gamma (k_z+1)\overline {n}_{\mathbf k}(\mu)\overline {n}_{\tilde{\mathbf k}}(\mu)+k_z\overline {n}_{\mathbf k}(\mu)]\times\\
              &\frac{\sin^2(v\hbar\omega_z/2)}{\hbar\omega_z}A(\omega_z)+\overline{N}(\mu)\left[-iv(\lambda_1^2-\lambda_0^2)+\frac{e^{iv\hbar\omega_z}-1}{\hbar\omega_z}A(\omega_z)\right],
  \end{split}
 \end{gather}
where $\overline{N}(\mu)=\sum_{\mathbf k}\overline {n}_{\mathbf k}(\mu)$ is the average particle number in the grand canonical ensemble. We would like emphasize that Eqs.~(\ref{e22}, \ref{e23}) are valid for both bosons and fermions and arbitrary large $N$, where previous methods fail~\cite{in2014}.

Based on the analytical solutions of the CFW (Eqs.~(\ref{e22}, \ref{e23})), we study the properties of the work statistics in several special cases:
\begin{enumerate}[fullwidth,itemindent=0em,label=(\arabic*)]
\item Single-particle case ($N=1$). In this case, we have $Z_1=\sum_{\mathbf k}e^{-\beta\varepsilon_{\mathbf k}}=[8\sinh(\beta\omega_x/2)\sinh(\beta\omega_y/2)\sinh(\beta\omega_z/2)]^{-1}$, $\overline {n}_{\mathbf k}(1)=e^{-\beta\varepsilon_{\mathbf k}}/Z_1$, $\xi_{\mathbf k}(1)=0$ (Eqs.~(\ref{e17}, \ref{e20})) and accordingly
 \begin{gather}
  \begin{split}
  \label{e21}
\ln \chi_1(v)=&\frac{-4\sin^2(v\hbar\omega_z/2)}{\hbar\omega_z(e^{\beta\hbar\omega_z}-1)}A(\omega_z)-iv(\lambda_1^2-\lambda_0^2)\\
&+\frac{e^{iv\hbar\omega_z}-1}{\hbar\omega_z}A(\omega_z).
  \end{split}
 \end{gather}
Actually, Eq.~(\ref{e21}) is identical to the exact expression of the cumulant CFW in Ref.~\cite{sta2008}, which indicates that for $N=1$, the contributions from the third or higher orders of $\lambda(s)$ vanish (see the supplemental material).
\item Nondegenerate case (in the thermodynamic limit $\varepsilon_0\sim\hbar\omega_z\sim N^{-1/3}$ and $k_B T\gg N^{1/3}\hbar\omega_{g}$).
From Eq.~(\ref{e21}), we obtain the classical limit of the cumulant CFW for a single particle~\cite{qu2018}
\be
\ln \chi_1^{\mathrm{cl}}(v)=-v^2k_B TA(\omega_z)-iv[\lambda_1^2-\lambda_0^2-A(\omega_z)].
\ee

In the nondegenerate case, we have $Z_1\approx(k_BT)^3/(\hbar\omega_g)^3$ which is equal to the partition function of a classical harmonic oscillator, $Z_N=Z_1^N/N!$, $Z_{\mu}=e^{\alpha Z_1}$,
$\overline {n}_{\mathbf k}(N)=Ne^{-\beta\varepsilon_{\mathbf k}}/Z_1$, $\overline {n}_{\mathbf k}(\mu)=\alpha e^{-\beta\varepsilon_{\mathbf k}}$, $\overline {n}_{\mathbf k}(\mu)\overline {n}_{\tilde{\mathbf k}}(\mu)\ll\overline {n}_{\mathbf k}(\mu)$, $\xi_{\mathbf k}(N)\ll\overline {n}_{\mathbf k}(N)$ (dilute gas).
From Eqs.~(\ref{e22}, \ref{e23}), we obtain the cumulant CFWs for many particles in two ensembles
\be
\ln \chi_N^{\mathrm{cl}}(v)=N\ln \chi_1^{\mathrm{cl}}(v),\ \ \ \ln \chi_{\mu}^{\mathrm{cl}}(v)=\alpha Z_1\ln \chi_1^{\mathrm{cl}}(v),
\ee
which indicates that the particles satisfy Maxwell-Boltzmann statistics and the contributions from the third or higher orders of $\lambda(s)$ vanish.
\end{enumerate}
The discussions about the CFW in the degenerate case 
is shown in the supplemental material.

\textit{Summary}.---The CFW is an important quantity to characterize the nonequilibrium process of the time-dependent quantum systems, especially of quantum many-body systems. But the calculation of the CFW for quantum many-body systems has been a long-time conundrum. To overcome this difficulty,  by utilizing the nonequilibrium Green's function's method, we formulate the work statistics with a contour integral and obtain the series expansion of the CFW. This method is valid for arbitrary $\hat H_0$ and $\hat H_1$, as well as for arbitrary perturbative work protocols $\lambda(s)$. Hence, it provides a unified method for the calculation of the CFW. In this framework, work, although not an observable, is defined as a functional along the modified contour. To the second order of the work parameter, the CFW is expressed as the sum of the first-order term, the ``speed'' term and the ``boundary'' term. Moreover, the fluctuation theorems can be demonstrated by utilizing the Kubo-Martin-Schwinger condition.
As an example, we calculate the CFW of noninteracting identical particles in a forced harmonic potential, where previous methods fail. In the future, we expect to investigate the effects of relativity and interactions with our methods.

H. T. Quan gratefully acknowledges support from
the National Science Foundation of China under grants
11775001, 11534002, and 11825001.

 \appendix
 \renewcommand{\theequation}{A.\arabic{equation}}

 \setcounter{equation}{0}

 \section*{APPENDIX A: The derivation of Eq.~(8) from Eq.~(4)}

In the following, we give the details about the derivation of Eq.~(8) from Eq.~(4). Following the same procedure as that in Eqs.~(7.21-7.23) in Ref.~\cite{th1952}, the series expansion of the cumulant CFW $\ln \chi(v)$ can be straightforwardly expressed as the integrals of the $n$-point cumulant correlation functions $G_c(s_1,\cdots,s_n)$. That is Eq.~(6) in the main text. Then, to the second order of the work parameter $\lambda(s)$, we have
\begin{widetext}
 \begin{gather}
  \begin{split}
\label{se6}
\ln\chi(v)=&\left(\int_{C'} \mathrm d\bar{s}_1-\int_{0}^{-i\hbar\beta}\mathrm d\bar{s}_1\right)G_c(s_1)+\left(\int_{C'} \mathrm d\bar{s}_1\int_{C'} \mathrm d\bar{s}_2-\int_{0}^{-i\hbar\beta}\mathrm d\bar{s}_1\int_{0}^{-i\hbar\beta}\mathrm d\bar{s}_2\right)G_c(s_1,s_2)+O(\lambda(s)^3),\\
=&\left(\int_{C'} \mathrm d\bar{s}_1-\int_{0}^{-i\hbar\beta}\mathrm d\bar{s}_1\right) G_c(s_1)+\left(\int_{C'} \mathrm d\bar{s}_1\int_{C'} \mathrm d\bar{s}_2-\int_{0}^{-i\hbar\beta}\mathrm d\bar{s}_1\int_{0}^{-i\hbar\beta}\mathrm d\bar{s}_2\right)G^>_c(s_1-s_2)+O(\lambda(s)^3),\\
=&iv(\lambda_1-\lambda_0)\langle\hat{H}_1\rangle_c+\int_{-\infty}^{\infty}\frac{\mathrm{d}\omega}{2\pi}G^>_c(\omega)\left(\int_{C'} \mathrm d\bar{s}_1\int_{C'} \mathrm d\bar{s}_2-\int_{0}^{-i\hbar\beta}\mathrm d\bar{s}_1\int_{0}^{-i\hbar\beta}\mathrm d\bar{s}_2\right)e^{-i\omega(s_1-s_2)}+O(\lambda(s)^3),
  \end{split}
 \end{gather}
 \end{widetext}
where $C'$ denotes the contour for work statistics (see Fig.~1b) in the main text. Since the work parameter $\lambda(s)$ is a piecewise function along the contour $C'$, we divide the interval of the integral along $C'$ into four parts:
\begin{enumerate}[fullwidth,itemindent=0em,label=(\roman*)]

\item Part 1, $s\in[0,t], \lambda_{C'}(s)=\lambda(s)$;
\item Part 2, $s\in[t,t-v], \lambda_{C'}(s)=\lambda_1$;
\item Part 3, $s\in[t-v,-v], \lambda_{C'}(s)=\lambda(s+v)$;
\item Part 4, $s\in[-v,-i\beta], \lambda_{C'}(s)=\lambda_0$.
\end{enumerate}
Then, the double integral along $C'$ in Eq.~\eqref{se6} is equal to the sum of the double integrals in Part $(i,j)$ ($i,j=1,2,3,4.$), i.e., the double integral along  $C'$ in Eq.~\eqref{se6} consists of 16 terms, and each term is labeled by a pair of $(i,j)$. Notice that due to the contour step function $\theta_{C'}(s_1-s_2)$ in $\mathrm d\bar{s}_1$, the double integrals for $i<j$ (6 terms) are equal to zero.
According to the value of the work parameter $\lambda(s)$ in four parts along the contour, we can further classify the 10 non-zero terms into 6 sets. For every set, we give the expression of the sum of  the double integral:
\begin{enumerate}[fullwidth,itemindent=0em,label=(\roman*)]
\item $(i,j)=(2,2)$:
\be
\frac{\lambda_1^2(1-e^{i\omega \hbar v}+i\omega \hbar v)}{\omega^2};
\ee
\item $(i,j)= (4,2)$:
\be
\frac{-2\lambda_0\lambda_1}{\omega^2}(1-e^{i\omega \hbar v})\cos(\omega t);
\ee
\item $(i,j)=(4,4)$:
 \begin{gather}
  \begin{split}
\frac{\lambda_0^2[e^{-\beta\hbar\omega}(1-e^{-i\omega \hbar v})-i\omega \hbar v]}{\omega^2}\\
+\int_{0}^{-i\hbar\beta}\mathrm d\bar{s}_1\int_{0}^{-i\hbar\beta}\mathrm d\bar{s}_2e^{-i\omega(s_1-s_2)};
  \end{split}
 \end{gather}
\item $(i,j)=(1,1), (3,3),  (3,1)$:
\be
(1-e^{i\omega \hbar v})\int_{0}^{t}\mathrm ds_1\int_{0}^{t}\mathrm ds_2\lambda(s_1)\lambda(s_2)e^{i\omega (s_1-s_2)};
\ee
\item $(i,j)=(2,1), (3,2)$:
\be
\frac{-2\lambda_1}{\omega}(1-e^{i\omega \hbar v})\int_{0}^{t}\mathrm ds\lambda(s)\sin[\omega (t-s)];
\ee
\item $(i,j)= (4,1),  (4,3)$:
\be
\frac{-2\lambda_0}{\omega}(1-e^{i\omega \hbar v})\int_{0}^{t}\mathrm ds\lambda(s)\sin(\omega s).
\ee
\end{enumerate}

The double integral along  $C'$ in Eq.~\eqref{se6} is equal to the sum of the above 6 expressions:
\begin{widetext}
 \begin{gather}
  \begin{split}
  \label{se10}
&\int_{C'} \mathrm d\bar{s}_1\int_{C'} \mathrm d\bar{s}_2e^{-i\omega(s_1-s_2)}-\int_{0}^{-i\hbar\beta}\mathrm d\bar{s}_1\int_{0}^{-i\hbar\beta}\mathrm d\bar{s}_2e^{-i\omega(s_1-s_2)}\\
=&\frac{1-e^{i\omega \hbar v}}{\omega^2}\left\{\lambda_1^2-2\lambda_0\lambda_1\cos(\omega t)+\lambda_0^2+\omega^2\int_{0}^{t}\mathrm ds_1\int_{0}^{t}\mathrm ds_2\lambda(s_1)\lambda(s_2)e^{i\omega (s_1-s_2)}-2\omega\lambda_1\int_{0}^{t}\mathrm ds\lambda(s)\sin[\omega (t-s)]\right.\\
&\left.-2\omega\lambda_0\int_{0}^{t}\mathrm ds\lambda(s)\sin(\omega s)\right\}+\frac{i\hbar v(\lambda_1^2-\lambda_0^2)}{\omega}+\frac{\lambda_0^2[e^{-\beta\hbar\omega}(1-e^{-i\omega \hbar v})-(1-e^{i\omega \hbar v})]}{\omega^2}\\
=&\frac{1-e^{i\omega \hbar v}}{\omega^2}\left|\lambda_1e^{i\omega t}-\lambda_0-i\omega\int_{0}^{t}\mathrm ds\lambda(s)e^{i\omega s}\right|^2+\frac{i\hbar v(\lambda_1^2-\lambda_0^2)}{\omega}+\frac{\lambda_0^2[e^{-\beta\hbar\omega}(1-e^{-i\omega \hbar v})-(1-e^{i\omega \hbar v})]}{\omega^2}\\
=&\frac{1-e^{i\omega \hbar v}}{\omega^2}\left|\int_{0}^{t}\mathrm ds \dot{\lambda}(s)e^{i\omega s}\right|^2+\frac{i\hbar v(\lambda_1^2-\lambda_0^2)}{\omega}+\frac{\lambda_0^2[e^{-\beta\hbar\omega}(1-e^{-i\omega \hbar v})-(1-e^{i\omega \hbar v})]}{\omega^2}\\
=&\frac{1-e^{i\omega \hbar v}}{\omega^2}A(\omega)+\frac{i\hbar v(\lambda_1^2-\lambda_0^2)}{\omega}+\frac{\lambda_0^2[e^{-\beta\hbar\omega}(1-e^{-i\omega \hbar v})-(1-e^{i\omega \hbar v})]}{\omega^2}.
  \end{split}
 \end{gather}
 \end{widetext}
Substituting Eq.~(\ref{se10}) into Eq.~(\ref{se6}) and using the Kubo-Martin-Schwinger condition, we finally obtain Eq.~(8) in the main text.

\section*{APPENDIX B: Exact expression of the CFW by perturbation expansion}

When Wick's theorem can be applied and $G_c(s_1,s_2)$ is represented by an arrow in connected Feynman diagrams, there must not be connected Feynman diagrams to the third or higher order of $\lambda(s)$. Hence, Eq.~(8) in the main text is the exact expression of the CFW now. One example is a forced harmonic oscillator~\cite{sta2008}, where the time-dependent Hamiltonian is
\be
\hat H(s)=\frac{\hat{p}^2}{2m}+\frac{1}{2}m\omega_0^2\hat{x}^2+\lambda(s)\omega_0\sqrt{2m}\hat {x}.
\ee
And the exact expression is shown in Eq.~(23). Another example is a driven quantum scalar field~\cite{wor2019}, where the time-dependent Hamiltonian in the Heisenberg picture is
\be
\hat H(s)=\frac{1}{2}\int\mathrm d^3x[\hat{\pi}^2+(\nabla \hat\phi)^2+m^2\hat \phi+2\lambda(s)F(x)\hat \phi].
\ee
Here, $\lambda(s)$ and $F(x)$ are called the switching and the smearing functions respectively. Then from Eq.~(8), the exact expression of the CFW reads
\begin{widetext}
\be
\label{es3}
\ln \chi(v)=\int\frac{\mathrm {d}^3p|\tilde{F}(\mathbf p)|^2}{(2\pi)^32\omega_{\mathbf p}^3}\left[\frac{-4\sin^2(v\omega_z/2)}{e^{\beta\omega_{\mathbf p}}-1}A(\omega_{\mathbf p})-iv\omega_{\mathbf p}(\lambda_1^2-\lambda_0^2)+(e^{iv\omega_{\mathbf p}}-1)A(\omega_{\mathbf p})\right].
\ee
\end{widetext}
where $\omega_{\mathbf p}=\sqrt{\mathbf{p}^2+m^2}$, $\tilde{F}(\mathbf p)=\int\mathrm d^3xF(x)e^{i\mathbf{p}\cdot\mathbf{x}}$ and we have set $\hbar=c=1$. We would like to emphasize that Eq.~(\ref{es3}) extends the results for a special protocol in Ref.~\cite{wor2019} to the results for an arbitrary driving protocol.

 \renewcommand{\theequation}{B.\arabic{equation}}

 \setcounter{equation}{0}

\section*{APPENDIX C: The CFW for noninteracting identical particles: degenerate case ($\overline{N}(\mu)\approx N$, $\hbar\omega_g,\hbar\omega_z\sim N^{-1/3}$)}

In this section, we only discuss the perturbation expansion of the cumulant CFW with the grand canonical initial state $\ln \chi_{\mu}(v)$ to the second order of $\lambda(s)$ for simplicity.
To replace the sum in Eq.~(22) by an integral, let us first introduce two types of density of states, $g_0(\varepsilon)=\sum_{\mathbf k}\delta(\varepsilon+\varepsilon_0-\varepsilon_{\mathbf k})=\varepsilon^2/[2(\hbar\omega_g)^3]$ and $g_1(\varepsilon)=\sum_{\mathbf k}k_z\delta(\varepsilon+\varepsilon_0-\varepsilon_{\mathbf k})=\varepsilon^3/[6\hbar\omega_z(\hbar\omega_g)^3]$. Notice that $g_0(\varepsilon)\ll g_1(\varepsilon)$ when $\hbar\omega_z\ll k_BT$.

Then for bosons, when the temperature is higher than the critical temperature of Bose-Einstein condensation, i.e., $k_B T\geq k_BT_c\sim N^{1/3}\hbar\omega_{g}$, we have  $\beta\varepsilon_0,\beta\hbar\omega_z\ll 1$.  Also according to Eq.~(22), we have
 \begin{widetext}
 \begin{gather}
  \begin{split}
  \label{es5}
  \ln \chi_{\mu}(v)\approx&-v^2\hbar\omega_z A(\omega_z)\int_{0}^{\infty}\mathrm d\varepsilon g_1(\varepsilon)\overline {n}^B_{\varepsilon}(\mu)[1+\overline {n}^B_{\varepsilon}(\mu)]-Niv[\lambda_1^2-\lambda_0^2-A(\omega_z)]\\
  =&N\{-v^2k_BTA(\omega_z)-iv[\lambda_1^2-\lambda_0^2-A(\omega_z)]\}\\
  =&N\ln\chi_1^{cl}(v),
  \end{split}
 \end{gather}
 \end{widetext}
where $\overline {n}^B_{\varepsilon}(\mu)=1/(\alpha^{-1}e^{\beta\varepsilon}-1)$ and
\be
\label{es6}
N=\int_{0}^{\infty}\mathrm d\varepsilon g_0(\varepsilon)\overline {n}^B_{\varepsilon}(\mu)=\left(\frac{k_BT}{\hbar\omega_g}\right)^3\mathrm{Li}_3(\alpha),
\ee
where $\mathrm{Li}_n(x)=\sum_{l=0}^{\infty}x^l/l^n$ is the polylogarithm function.
When $\hbar\omega_z\ll k_B T<k_BT_c$, the contribution in the sum in Eq.~(22) from the particles in the single-particle ground state can not be ignored. And we have $\mu\approx\varepsilon_0$, $\overline {n}^{B}_{\hbar\omega_z}(\mu)\approx (k_BT)/(\hbar\omega_z)$, $\beta\varepsilon_0,\beta\hbar\omega_z\ll 1$. Thus according to Eq.~(22), we have
\begin{widetext}
 \begin{gather}
  \begin{split}
  \label{es7}
  \ln \chi_{\mu}(v)\approx&-v^2\hbar\omega_z A(\omega_z)\left\{\overline {n}^{B}_{0}(\mu)\overline {n}^{B}_{\hbar\omega_z}(\mu)+\int_{0}^{\infty}\mathrm d\varepsilon g_1(\varepsilon)\overline {n}^B_{\varepsilon}(\mu)[1+\overline {n}^B_{\varepsilon}(\mu)]\right\}-Niv[\lambda_1^2-\lambda_0^2-A(\omega_z)]\\
  =&N\{-v^2k_BTA(\omega_z)-iv[\lambda_1^2-\lambda_0^2-A(\omega_z)]\}\\
  =&N\ln\chi_1^{cl}(v),
  \end{split}
 \end{gather}
 \end{widetext}
where
\be
\overline {n}^{B}_{0}(\mu)=\frac{\alpha}{1-\alpha}=\left[1-\left(\frac{T}{T_c}\right)^3\right]N.
\ee
Finally when $k_BT\lesssim\hbar\omega_z\approx0$,  all particles are almost in the single-particle ground state and $\hbar\omega_z$ cannot be considered as a perturbation anymore. Almost all contributions in the sum in Eq.~(22) are from the particles in the single-particle ground state. We have $\mu\approx \varepsilon_0$, $N\approx\overline {n}^{B}_{0}(\mu)$, $\overline {n}^{B}_{\hbar\omega_z}(\mu)\approx 1/(e^{\beta\hbar\omega_z}-1)$. Thus according to Eq.~(22), we have
 \begin{widetext}
 \begin{gather}
  \begin{split}
  \label{es8}
  \ln \chi_{\mu}(v)\approx&  \frac{-4\sin^2(v\hbar\omega_z/2)}{\hbar\omega_z}A(\omega_z)\overline {n}^{B}_{0}(\mu)\overline {n}^{B}_{\hbar\omega_z}(\mu)+N\left[-iv(\lambda_1^2-\lambda_0^2)+\frac{e^{iv\hbar\omega_z}-1}{\hbar\omega_z}A(\omega_z)\right]\\
  =&N\ln \chi_1(v).
  \end{split}
 \end{gather}
\end{widetext}

For fermions, when $\hbar\omega_z\ll k_B T$, 
 we have $\beta\varepsilon_0,\beta\hbar\omega_z\ll 1$. Thus according to Eq.~(22), we have
 \begin{widetext}
 \begin{gather}
  \begin{split}
  \label{es9}
  \ln \chi_{\mu}(v)\approx&-v^2\hbar\omega_z A(\omega_z)\int_{0}^{\infty}\mathrm d\varepsilon g_1(\varepsilon)\overline {n}^F_{\varepsilon}(\mu)[1-\overline {n}^F_{\varepsilon}(\mu)]-Niv[\lambda_1^2-\lambda_0^2-A(\omega_z)]\\
  =&N\{-v^2k_BTA(\omega_z)-iv[\lambda_1^2-\lambda_0^2-A(\omega_z)]\}\\
  =&N\ln\chi_1^{cl}(v),
  \end{split}
 \end{gather}
 \end{widetext}
where $\overline {n}^F_{\varepsilon}(\mu)=1/(\alpha^{-1}e^{\beta\varepsilon}+1)$ and
\be
N=\int_{0}^{\infty}\mathrm d\varepsilon g_0(\varepsilon)\overline {n}^F_{\varepsilon}(\mu)=-\left(\frac{k_BT}{\hbar\omega_g}\right)^3\mathrm{Li}_3(-\alpha).
\ee
When $k_BT\lesssim\hbar\omega_z\approx0$, $\hbar\omega_z$ cannot be considered as a perturbation anymore but $\varepsilon_0$ can still be ignored due to the large $\mu$. Then according to Eq.~(22), we have
 \begin{widetext}
 \begin{gather}
  \begin{split}
  \label{es10}
\ln \chi_{\mu}(v)\approx&\frac{-4\sin^2(v\hbar\omega_z/2)}{\hbar\omega_z}A(\omega_z)\int_{0}^{\infty}\mathrm d\varepsilon \left\{g_1(\varepsilon)\overline {n}^F_{\varepsilon}(\mu)[1-\overline {n}^F_{\varepsilon+\hbar\omega_z}(\mu)]-g_0(\varepsilon)\overline {n}^F_{\varepsilon}(\mu)\overline {n}^F_{\varepsilon+\hbar\omega_z}(\mu)\right\}\\
&+N\left[-iv(\lambda_1^2-\lambda_0^2)+\frac{e^{iv\hbar\omega_z}-1}{\hbar\omega_z}A(\omega_z)\right]\\
=&\frac{-4\sin^2(v\hbar\omega_z/2)}{\hbar\omega_z}A(\omega_z)\left(\frac{k_BT}{\hbar\omega_g}\right)^3\left\{\frac{e^{\beta\hbar\omega_z}[\mathrm{Li}_4(-\alpha e^{-\beta\hbar\omega_z})-\mathrm{Li}_4(-\alpha)]}{\beta\hbar\omega_z(e^{\beta\hbar\omega_z}-1)}+\frac{e^{\beta\hbar\omega_z}\mathrm{Li}_3(-\alpha e^{-\beta\hbar\omega_z})-\mathrm{Li}_3(-\alpha)}{e^{\beta\hbar\omega_z}-1}\right\}\\
&+N\left[-iv(\lambda_1^2-\lambda_0^2)+\frac{e^{iv\hbar\omega_z}-1}{\hbar\omega_z}A(\omega_z)\right]\\
\approx& N\ln \chi_1(v),
  \end{split}
 \end{gather}
 \end{widetext}
where $N=\mu^3/6(\hbar\omega_g)^3$. Here in the calculation, we have used the property: for large $\alpha$, $-\mathrm{Li}_3(-\alpha)\approx(\ln\alpha)^3/3!$.

From the above analysis, we found that: (1) the cumulant CFW for the degenerate case is approximately equal to that of a single particle multiplied by a factor $N$; (2) When $k_BT\gg\hbar\omega_z$, the cumulant CFW for the single particle is replaced by its classical counterpart. We would like to emphasize that the multiplicity relation between the many-particle system and a single-particle system is due to the peculiarity of this model. For a generic model, e.g., a harmonic potential with a time-dependent frequency, the cumulant  CFW of a many-particle system is not equal to that of a single particle multiplied by a factor $N$.


\begin{thebibliography}{99}

\bibitem{st2010} K. Sekimoto, \emph{Stochastic energetics} (Springer 2010).
\bibitem{eq2011} C. Jarzynski, Annu. Rev. Condens. Matter Phys. \textbf{2}(1), 329-351 (2011).
\bibitem{st2012} U. Seifert, Reports on progress in physics \textbf{75}(12), 126001 (2012).
\bibitem{non2013} R. Klages, W. Just, and C. Jarzynski, (Eds.) \emph{Nonequilibrium statistical physics of small systems.} (Wiley-VCH Verlag GmbH and Company KGaA, 2013).
\bibitem{ge2012} B. D\'{o}ra, \'{A}. B\'{a}csi, and G. Zar\'{a}nd, Phys. Rev. B \textbf{86}(16), 161109(R) (2012).
\bibitem{th2018} J. Goold, F. Plastina, A. Gambassi, and A. Silva, arxiv: 1804.02805 (2018).
\bibitem{as2015} A. Russomanno, S. Sharma, A. Dutta and G. E Santoro, J. Stat. Mech, P08030 (2015).
\bibitem{wor2019} A. Ortega, E. McKay, \'{A}. M. Alhambra and E. Mart\'{i}n-Mart\'{i}nez,  Phys. Rev. Lett. \textbf{122}(24), 240604 (2019).
\bibitem{aq2000} J. Kurchan, arXiv preprint cond-mat/0007360 (2000).
\bibitem{ja2000} H. Tasaki, arXiv preprint cond-mat/0009244 (2000).
\bibitem{flu2007} P. Talkner, E. Lutz and P. H\"{a}nggi, Phys. Rev. E \textbf{75}, 050102(R) (2007).
\bibitem{st2008} A. Silva, Phys. Rev. Lett. \textbf{101}, 120603 (2008).
\bibitem{de2006} H. T. Quan, Z. Song, X. F. Liu, P. Zanardi, and C. P. Sun, Phys. Rev. Lett. \textbf{96}(14), 140604 (2006).
\bibitem{fzy2020} Z. Y. Fei, N. Freitas, V. Cavina, H. T. Quan, and M. Esposito, Phys. Rev. Lett. \textbf{124}(17), 170603 (2020).
\bibitem{dy2013} M. Heyl, A. Polkovnikov, and S. Kehrein, Phys. Rev. Lett. \textbf{110}, 135704 (2013).
\bibitem{qu2016} N. O. Abeling and S. Kehrein, Phys. Rev. B \textbf{93}, 104302 (2016).
\bibitem{sta2008} P. Talkner, P. S. Burada and P. H\"{a}nggi, Phys. Rev. E \textbf{78}, 011115 (2008).
\bibitem{qu2017} J. D. Jaramillo, J. Deng, and J. Gong, Phys. Rev. E \textbf{96}(4), 042119 (2017).
\bibitem{sta2019} H. K. Yadalam and U. Harbola, Phys. Rev. A \textbf{99}, 063802 (2019).
\bibitem{no2008} S. Deffner, and E. Lutz, Phys. Rev. E \textbf{77}, 021128 (2008).
\bibitem{wo2013} P. Smacchia and A. Silva, Phys. Rev. E \textbf{88}(4), 042109 (2013).

\bibitem{em2012} R. Dorner, J. Goold, C. Cormick, M. Paternostro and V. Vedral, Phys. Rev. Lett. \textbf{109}, 160601 (2012).
\bibitem{fu2019} J. J. Dong, and Y. F. Yang, Phys. Rev. B \textbf{100}, 035124 (2019).
\bibitem{wo2019} E. G. Arrais, D. A. Wisniacki, A. J. Roncaglia, and F. Toscano, arxiv: 1907.06285 (2019).
\bibitem{qu2019} Z. Y. Fei, J. N. Zhang, R. Pan, T. Qiu and H. T. Quan, Phys. Rev. A \textbf{99}, 052508 (2019).
\bibitem{ja2018} A. Bartolotta, and S. Deffner, Phys. Rev. X \textbf{8}(1), 011033 (2018).

\bibitem{gr2019} Z. Y. Fei,  and H. T. Quan, Phys. Rev. Research, \textbf{1}(3), 033175 (2019).
\bibitem{no2013} K. Balzer and M. Bonitz, \emph{Nonequilibrium Green's Functions Approach to Inhomogeneous Systems.} (Spring, 2013).
\bibitem{di1965} L. V. Keldysh, Sov. Phys. JETP, \textbf{20}(4), 1018-1026 (1965).
\bibitem{la1992} Y. Meir, and N. S.  Wingreen,  Phys. Rev. Lett. \textbf{68}(16), 2512 (1992).
\bibitem{fl2007} K. Saito and A. Dhar, Phys. Rev. Lett.  \textbf{99}, 180601 (2007).
\bibitem{no2018} C. Aron, G. Biroli, and L. F. Cugliandolo, SciPost Phys. \textbf{4}(1), 008, (2018).
\bibitem{sy2019} J. Yeo, Phys. Rev. E \textbf{100}, 062107 (2019).
\bibitem{ex2013} R. Dorner, S. R. Clark, L. Heaney, R. Fazio, J. Goold, and V. Vedral, Phys. Rev. Lett. \textbf{110}, 230601 (2013).
\bibitem{pa2018} K. Funo and H. T. Quan, Phys. Rev. Lett. \textbf{121}, 040602 (2018).
\bibitem{no1997} C. Jarzynski, Phys. Rev. Lett. \textbf{78}(14), 2690-2693 (1997).

\bibitem{qu1987} J. Glimm, and A. Jaffe, \emph{Quantum physics: a functional integral point of view. 2ed edition} (Springer, 2012).
\bibitem{th1952} R. Kubo, Journal of the Physical Society of Japan, \textbf{17}(7), 1100-1120 (1962).
\bibitem{gr2006} E. N. Economou,  \emph{Green's functions in quantum physics.} (Springer Science and Business Media, 2006).
\bibitem{foot1} Actually, this definition is correct only when $\hat H_1$ is even under the time reversal. When $\hat H_1$ is odd under the time reversal, we should define $\hat H_B(s)=\hat H_0-\lambda(t-s)\hat H_1$~\cite{th2009}.

\bibitem{th2009} D. Andrieux, P. Gaspard, T. Monnai, and S. Tasaki, New J. Phys. \textbf{11}, 043014 (2009).
\bibitem{on1951}  J. Schwinger, Phys. Rev. \textbf{82}(5), 664-679 (1951).

\bibitem{en1999} G. E. Crooks, Phys. Rev. E \text{60}(3), 2721-2726 (1999).

\bibitem{st2011} R. K. Pathria and P. D. Beale, \emph{Statistical Mechanics. 3rd edition} (Elsevier, 2011).
\bibitem{th1999} F. Dalfovo, S. Giorgini, L. P. Pitaevskii and S. Stringari, Rev. Mod. Phys. \textbf{71}(3), 463-512 (1999).

\bibitem{in2014} Z. Gong, S. Deffner, and H. T. Quan, Phys. Rev. E \textbf{90}(6), 062121 (2014).
\bibitem{qu2018} Z. Y. Fei, H. T. Quan and F. Liu, Phys. Rev. E \textbf{98}, 012132 (2018).





\end{thebibliography}
\end{document}